\pdfoutput=1

\documentclass[final,1p,times]{elsarticle}

\usepackage{bbold}
\usepackage{mathrsfs}
\usepackage{amsmath}
\usepackage[colorlinks]{hyperref}

\journal{Nuclear Physics B}
\biboptions{sort&compress}
\allowdisplaybreaks

\let\a=\alpha
\let\b=\beta
\let\g=\gamma
\let\d=\delta

\let\m=\mu
\let\n=\nu
\let\o=\omega
\let\t=\theta
\newcommand{\inv}{^{-1}}
\newcommand{\im}{\text{i}}
\newcommand{\dd}{\text{d}}
\newcommand{\gr}[1]{\text{#1}}
\newcommand\Aa{\mathcal{A}}
\newcommand\La{\mathscr{L}}
\renewcommand\P{\mathcal{P}}
\newcommand\R{\mathcal{R}}
\newcommand\RA{A_\R}
\newcommand\RF{F_\R}
\newcommand\he[1]{{#1}^\dagger}
\DeclareMathOperator{\tr}{tr}

\begin{document}
 
\begin{frontmatter}

\title{Gauged Wess-Zumino terms for a general coset space}
\author{Tom\'a\v{s} Brauner}
\ead{tomas.brauner@uis.no}
\author{Helena Kole\v{s}ov\'{a}}
\ead{helena.kolesova@uis.no}

\address{Department of Mathematics and Physics, University of Stavanger, N-4036 Stavanger, Norway}

\begin{abstract}
The low-energy physics of systems with spontaneously broken continuous symmetry is dominated by the ensuing Nambu-Goldstone bosons. It has been known for half a century how to construct invariant Lagrangian densities for the low-energy effective theory of Nambu-Goldstone bosons. Contributions, invariant only up to a surface term -- also known as the Wess-Zumino (WZ) terms -- are more subtle, and as a rule are topological in nature. Although WZ terms have been studied intensively in theoretically oriented literature, explicit expressions do not seem to be available in sufficient generality in a form suitable for practical applications. Here we construct the WZ terms in $d=1,2,3,4$ spacetime dimensions for an arbitrary compact, semisimple and simply connected symmetry group $G$ and its arbitrary connected unbroken subgroup $H$, provided that the $d$-th homotopy group of the coset space $G/H$ is trivial. Coupling to gauge fields for the whole group $G$ is included throughout the construction. We list a number of explicit matrix expressions for the WZ terms in four spacetime dimensions, including those for QCD-like theories, that is vector-like gauge theories with fermions in a complex, real or pseudoreal representation of the gauge group.
\end{abstract}


\end{frontmatter}


\section{Introduction}
\label{sec:intro}

Spontaneous breaking of a continuous global symmetry is a fascinating phenomenon that dramatically affects the physics in the quantum realm. Its most striking consequence is the presence of gapless modes --- the Nambu-Goldstone (NG) bosons --- in the spectrum which, in absence of other gapless degrees of freedom, dominate the low-energy physics of the system. Another, related consequence is the fact that spontaneously broken symmetry is not realized by unitary operators on the Hilbert space of physical states of the system. By the same token, transformations induced by the spontaneously broken generators of the symmetry group $G$ are realized nontrivially even on the space of fields, through nonlinear functions of the NG fields.

The latter feature presents a challenge to model building. In general, the \emph{action} of the low-energy effective theory, formulated in terms of the NG degrees of freedom, must preserve the symmetry properties of the underlying microscopic theory~\cite{Leutwyler:1993iq,Leutwyler:1993gf}. The mathematical problem of how to build $G$-invariant \emph{Lagrangians} was resolved, at least for compact groups $G$, by Callan et al.~through the so-called coset construction~\cite{Coleman:1969sm,Callan:1969sn}. However, invariance of the action does not necessarily imply invariance of the Lagrangian density. It is sufficient that the latter changes upon a symmetry transformation by a surface term, in which case we speak of a quasi-invariant Lagrangian.

Quasi-invariant Lagrangians appeared first in high-energy physics in connection to chiral anomalies through the work of Wess and Zumino~\cite{Wess:1971yu}. Accordingly, we will refer to them equivalently as Wess-Zumino (WZ) terms. However, it turns out that the correspondence between WZ terms and anomalies only holds in an even number of spacetime dimensions. In an odd number of spacetime dimensions, there are no chiral anomalies, yet WZ terms can still be present~\cite{Brauner:2014ata}; see ref.~\cite{Altland:book} for a self-contained non-technical introduction to WZ terms in the context of condensed-matter physics.

It was first shown by Witten~\cite{Witten:1983tw} that the anomalous WZ term in quantum chromodynamics (QCD) in four spacetime dimensions can be obtained from a strictly invariant Lagrangian in a five-dimensional spacetime. That a similar relation holds in general was confirmed by an analysis of quasi-invariant Lagrangians in ref.~\cite{DHoker:1994ti}. Its authors proved that, with certain technical assumptions on the topology of the coset space $G/H$, where $H$ denotes the unbroken subgroup, WZ terms can be classified in terms of the de Rham cohomology of $G/H$. Thus, WZ terms in $d$ spacetime dimensions correspond to invariant Lagrangians in $d+1$ dimensions that cannot be reduced to a $d$-dimensional invariant Lagrangian density. All the cohomology generators up to degree 5, or equivalently all quasi-invariant Lagrangians in spacetime dimension up to four, were constructed explicitly in ref.~\cite{DHoker:1995it} for an arbitrary compact, semisimple and simply connected $G$ and its arbitrary connected subgroup $H$.

Numerous physical applications require understanding of not only the dynamics of the NG bosons, but also of their interactions with external fields or of the correlation functions of the corresponding broken Noether currents. Such questions are most conveniently addressed by coupling the action to a set of gauge fields for the generators of the group $G$. It is known that in odd-dimensional spacetimes, gauging of the WZ terms is straightforward, whereas in even-dimensional spacetimes, it faces a geometric obstruction, related to the chiral anomaly. In general, only the unbroken subgroup $H$ can be gauged without violating the invariance of the action.

In spite of abundant literature on the subject, a sufficiently general expression for gauged WZ terms that would at the same time be explicit enough and ready-made for applications, seems to be missing. Existing results for gauged WZ terms are in most cases either limited to chiral symmetry groups of the type found in QCD (see, for instance, refs.~\cite{Kaymakcalan:1983qq,Manohar:1984uq,AlvarezGaume:1984dr,Manes:1984gk}), or presented in an implicit form that requires a substantial amount of mathematical formalism to convert to a form suitable for applications~\cite{Hull:1990ms,Wu1993381,Figueroa-OFarrill:1994vwl}.

The aim of the present paper is to fill this gap in the literature. In order to keep the text concise, we do not provide any computational details; the methodology is explained in ref.~\cite{DHoker:1995it} that we largely follow. The necessary minimum of formalism and notation are reviewed in section~\ref{sec:notation}, the rest of the paper is devoted to the summary of our results. The general results for WZ terms in different spacetime dimensions are listed in section~\ref{sec:results}, and in section~\ref{sec:examples} they are adapted for application to several systems of physical interest.

We do \emph{not} spell out our results as explicit Lagrangians, but rather using the language of differential forms which can then be easily translated, see section~\ref{sec:formstolag} for a basic dictionary. In addition to the obvious advantage of making the expressions more compact, this also underlines their multi-faceted use. For instance, the same WZ 3-form can describe a topological action in two spacetime dimensions and a topological current in four spacetime dimensions. Apart from the moderate use of differential forms, the only mathematical background required for utilizing our results for practical calculations is some standard Lie algebra and group theory and rudiments of topology, along with the usual coset construction of effective Lagrangians.

Following the analysis of ref.~\cite{DHoker:1995it}, the list of gauged WZ terms provided here is exhaustive for cohomologies of degrees 2 and 4, and as far as gauging of the unbroken subgroup $H$ is concerned for degrees 3 and 5. While we make no claim regarding the uniqueness of the \emph{fully} gauged cohomology generators of degrees 3 and 5, we point out that the results given here naturally recover the known chiral anomalies in two and four spacetime dimensions.


\section{Notation and formalism}
\label{sec:notation}


\subsection{Wess-Zumino terms}\label{subsec:WZterms}

The starting point of Witten's construction of the WZ terms is the assumption that the NG fields of the theory, represented collectively by the matrix $U(x)$, satisfy a boundary condition that effectively allows the $d$-dimensional spacetime to be compactified to the sphere $S^d$. Suppose in addition that the $d$-th homotopy group of the coset space $G/H$ is trivial, $\pi_d(G/H)=0$.\footnote{See ref.~\cite{Davighi:2018inx} for a recent classification of topological terms which avoids assumptions on the homotopy of the coset space.} The latter condition ensures that one can always find an interpolating field $\tilde U_t(x)$ with $t\in[0,1]$ such that $\tilde U_1(x)=U(x)$ and $\tilde U_0(x)$ is constant. In other words, the image of spacetime in $G/H$ defined by the NG fields, $U(S^d)$, can always be contracted to a point. As $t$ varies between $0$ and $1$, the field $\tilde U_t(x)$ sweeps a $(d+1)$-dimensional disc $D^{d+1}$ in $G/H$ with the boundary $U(S^d)$. Physically well-defined invariant actions that are local functionals of the NG field $U(x)$, independent of the extension $\tilde U_t(x)$, can be represented by a closed invariant $(d+1)$-form, $\o_{d+1}$~\cite{DHoker:1994ti}. Being closed, this form can be locally written as a derivative of a $d$-form, $\tilde\o_d$. By the Stokes theorem, we thus get the WZ action
\begin{equation}
S_\text{WZ}\propto \int_{D^{d+1}}\o_{d+1}=\int_{D^{d+1}}\dd\tilde\o_d=\int_{U(S^d)}\tilde\o_d=\int_{S^d}U^*\tilde\o_d.
\label{master}
\end{equation}
Accordingly, the pull-back $U^*\tilde\o_d$ gives the corresponding $d$-dimensional Lagrangian density. If $\tilde\o_d$ itself is $G$-invariant, then $\o_{d+1}$ is also necessarily $G$-invariant. Nontrivial WZ terms therefore stem from such closed forms $\o_{d+1}$ that are $G$-invariant, whereas the corresponding $\tilde\o_d$ is not. These forms define precisely the cohomology of the coset space $G/H$.

In section~\ref{sec:results} below, we outline the classification of the cohomology generators $\o_{d+1}$ up to degree 5. The more physical $d$-forms $\tilde\o_d$ can be deduced therefrom by means of eq.~\eqref{master}. The most important new result, obtained in this paper, is an explicit general expression for the gauge-field-dependent part of $\tilde\o_4$, corresponding to quasi-invariant effective Lagrangians in four spacetime dimensions.

As indicated by the proportionality symbol in eq.~\eqref{master}, the overall normalization of the form $\o_{d+1}$ is not fixed and the WZ action is accordingly defined only up to a multiplicative factor, which has to be determined by matching to an underlying microscopic theory. In case that the homotopy group $\pi_{d+1}(G/H)$ is nontrivial, the independence of the physical action on the choice of the reference point $\tilde U_0(x)$ requires this factor to be quantized~\cite{Witten:1983tw}. This topological constraint is absent when $\pi_{d+1}(G/H)$ is trivial, see ref.~\cite{Bar:2004bw} for an example.


\subsection{Gauged Maurer-Cartan 1-form}

Following largely the notation of ref.~\cite{DHoker:1995it}, we shall denote all the generators of $G$ collectively as $T_{A,B,\dotsc}$ and fix the structure of the Lie algebra they span by
\begin{equation}
[T_A,T_B]=f_{AB}^{\phantom{AB}C}T_C.
\label{fABC}
\end{equation}
The generators of the unbroken subgroup $H$ will be denoted as $T_{\a,\b,\dotsc}$, whereas the broken generators as $T_{a,b,\dotsc}$. According to the coset construction~\cite{Coleman:1969sm,Callan:1969sn}, the matrix NG field $U(x)$ transforms under an element $g\in G$ as
\begin{equation}
U\xrightarrow{g}U'=gUh\inv,
\label{coset}
\end{equation}
where $g\in G$ and $h\in H$. We will right away consider local $g$, that is, assume that the symmetry is gauged. Although $h$ in eq.~\eqref{coset} depends in general on both $g$ and $U$, it is convenient to imagine that the matrices $g$ and $h$ are completely independent, that is, to picture the transformation rule~\eqref{coset} as defining the action of the group $G\times H$, whereby $G$ acts on $U$ from the left and $H$ from the right.

As the next step, we introduce the Lie-algebra-valued Maurer-Cartan (MC) 1-form,
\begin{equation}\label{def:bart}
\bar\t\equiv\bar\t^AT_A\equiv U\inv(\dd+A)U=\t+\bar A,
\end{equation}
where $\t\equiv U\inv\dd U$ and $\bar A\equiv U\inv AU$. Herein, $A$ is the 1-form gauge connection of $G$, which under a $g\in G$ transforms as
\begin{equation}
A\xrightarrow{g}gAg\inv+g\dd g\inv.
\end{equation}
It follows that the MC form is invariant under the left action of $G$ on $U$. Under the right action of $H$, it transforms as $\bar\t\xrightarrow{h}h\bar\t h\inv+h\dd h\inv$. Since the second term on the right-hand side of this transformation rule belongs to the Lie algebra of $H$, it is convenient to split the MC form into the broken and unbroken components,
\begin{equation}
\bar\phi\equiv\bar\t^aT_a\xrightarrow{h}h\bar\phi h\inv,\qquad
\bar V\equiv\bar\t^\a T_\a\xrightarrow{h}h\bar Vh\inv+h\dd h\inv.
\end{equation}
The $V$-part behaves as a gauge connection of $H$ and gives rise to the field-strength 2-form\footnote{We mostly omit the $\wedge$ symbol in products of forms. Thus, for instance, $\bar V^2$ is a shorthand notation for $T_\a T_\b\bar\t^\a\wedge\bar\t^\b=\frac12[T_\a,T_\b]\bar\t^\a\wedge\bar\t^\b=\frac12f_{\a\b}^{\phantom{\a\b}\g}T_\g\bar\t^\a\wedge\bar\t^\b$.}
\begin{equation}
\bar W\equiv\dd\bar V+\bar V^2\xrightarrow{h}h\bar Wh\inv.
\end{equation}
While the gauged MC form is essential for the general construction of WZ terms, the ungauged versions of the above-defined objects will also be needed. These are naturally defined by
\begin{equation}
\phi\equiv\t^aT_a,\qquad
V\equiv\t^\a T_\a,\qquad
W\equiv\dd V+V^2.
\end{equation}
The gauge connection of $G$ gives rise to the field-strength 2-form
\begin{equation}
F\equiv\dd A+A^2\xrightarrow{g}gFg\inv.
\end{equation}
It is easy to check that the MC form $\bar\t$ satisfies the MC structure equation
\begin{equation}
\dd\bar\t+\bar\t^2=U\inv FU\equiv\bar F,\qquad
\dd\bar\t^A+\tfrac12f^A_{\phantom ABC}\bar\t^B\bar\t^C=\bar F^A.
\label{MCeq}
\end{equation}
Invariant Lagrangians can now be built out of the covariant building blocks $\bar\phi$, $\bar W$ and $\bar F$, which are all $G$-invariant and transform linearly under the adjoint action of $H$. Their covariant derivatives can be constructed using the gauge connection $\bar V$. Left $G$-invariance of the Lagrangian follows trivially from the $G$-invariance of all the basic building blocks. Right $H$-invariance, on the other hand, imposes nontrivial constraints. In general, it requires that the indices of the building blocks of a given operator be contracted by a coefficient that is an invariant tensor under the adjoint action of $H$.


\subsection{Recovering the Lagrangian form of WZ terms}
\label{sec:formstolag}

As already mentioned, our main results are presented as the differential forms $\omega_{d+1}$ and $\tilde{\omega}_d$ in eq.~\eqref{master}. For the sake of concrete applications, it may be more desirable to have the corresponding Lagrangian density $U^*\tilde\o_d$ at hand. This can be recovered from the $d$-form $\tilde\o_d$ by the following replacements,
\begin{equation}
A\to A_\m\dd x^\m,\quad
F\to\tfrac12F_{\m\n}\dd x^\m\dd x^\n,\quad
\dd\to\dd x^\m \partial_\m,\quad
\dd x^\m\dd x^\n\dd x^\rho\dotsb\to\epsilon^{\m\n\rho\dotsb}\dd^dx,
\end{equation}
where in the last relation $\epsilon^{\mu\nu\rho\ldots}$ stands for the Levi-Civita tensor in $d$ dimensions.

Note also that our fundamental commutation relation~\eqref{fABC} does not contain a factor of $\im$ on the right-hand side, as would be appropriate for compact Lie algebras. If desired, this can be recovered by the replacement $A\to-\im A$ in the final results written in terms of the matrix NG field $\Sigma$ and the gauge field $A$, see section~\ref{subsec:symmetric}.


\section{Results}
\label{sec:results}

Having put together all the required notation, we will now simply list all the independent cohomology generators of degrees 2 to 5 that are primitive, that is, cannot be obtained as a product of generators of lower degree. These accordingly generate all the independent WZ terms in spacetime dimension one to four via eq.~\eqref{master}. The case of most physical interest is the cohomology of degree 5. The cohomologies of lower degrees are of interest on their own though. First, they may be needed for the construction of non-primitive generators of the degree-5 cohomology. Second, they are of relevance for low-dimensional or nonrelativistic systems. 

Finally, note that in case the group $G$ is actually not gauged, the corresponding ungauged WZ terms can be recovered by simply setting $A\to 0$ and $F\to 0$ everywhere. There are no additional WZ terms in the ungauged case.


\subsection{Degree 2 (one dimension)}
\label{subsec:deg2}

The nontrivial cohomology generators of degree 2 can be parameterized by a set of constants $c_\a$, which are required to be invariant under the adjoint action of $H$,
\begin{equation}
c_\a f^\a_{\phantom\a\b\g}=0.
\label{Hinv}
\end{equation}
For compact Lie groups, there is one such a constant for every $\gr{U(1)}$ factor of $H$. The corresponding closed gauge-invariant 2-form reads
\begin{equation}
\o_2=-c_\a\bar W^\a=\tfrac12c_\a f^\a_{\phantom\a bc}\bar\t^b\bar\t^c-c_\a\bar F^\a.
\label{deg2}
\end{equation}
In this case, one can readily obtain the corresponding 1-form, which directly gives the WZ contribution to the effective Lagrangian in one dimension,
\begin{equation}
\tilde\o_1=-c_\a\bar\t^\a.
\label{CS1}
\end{equation}
This result appeared previously in ref.~\cite{DHoker:1995it} and was rederived using elementary field theory in refs.~\cite{Brauner:2014ata,Watanabe:2014fva}. The Lagrangian~\eqref{CS1} describes for instance the dynamics of a single spin degree of freedom in an external magnetic field. It is also important for the dynamics of spin waves in ferromagnets~~\cite{Leutwyler:1993gf,Volkov:1971vo}. Its topological nature reflects the Berry phase acquired by the spin state when dragged by an external field around a closed loop in the group space.


\subsection{Degree 3 (two dimensions)}
\label{subsec:deg3}

In this case, the nontrivial cohomology generators are associated with constant symmetric $G$-invariant tensors $d_{AB}$ that vanish on the unbroken subgroup, that is, satisfy $d_{\a\b}=0$. The corresponding gauge-invariant 3-form reads
\begin{equation}
\o_3=\bigl(\tfrac16d_{ad}f_{bc}^{\phantom{bc}d}+\tfrac23d_{a\d}f_{bc}^{\phantom{bc}\d}\bigr)\bar\t^a\bar\t^b\bar\t^c-(d_{ab}\bar F^a\bar\t^b+2d_{\a b}\bar F^\a\bar\t^b).
\label{deg3aux}
\end{equation}
This form is, however, in general not closed, but rather satisfies
\begin{equation}
\dd\o_3=-d_{AB}F^AF^B.
\label{ddo3}
\end{equation}
The fact that the derivative $\dd\o_3$ only depends on the background gauge field indicates that in its absence, $\o_3$ generates a well-defined WZ term. The symmetry can, however, not be fully gauged. In other words, the WZ term in two spacetime dimensions is anomalous. Thanks to $d_{\a\b}=0$, it is nevertheless possible to gauge at least the unbroken subgroup $H$ while preserving closedness of $\o_3$.

To gain deeper insight into the anomalous nature of the WZ 3-form $\o_3$, it is convenient to switch to a matrix notation. Recall that simple compact Lie groups have a unique $G$-invariant symmetric tensor $d_{AB}$ given by the Cartan-Killing form on their Lie algebra. The most general invariant tensor $d_{AB}$ for a semisimple group $G$ can then be expressed as $d_{AB}=\sum_jd_j\tr_j(T_AT_B)$, where the sum runs over all simple factors of $G$, the trace $\tr_j$ is done over the $j$-th simple component of $G$, and $d_j$ is a set of coefficients only constrained by the requirement that $d_{\a\b}=0$. Introducing the shorthand notation
\begin{equation}
\langle X\rangle\equiv\sum_jd_j\tr_jX,
\label{tracenotation}
\end{equation}
we can then rewrite eqs.~\eqref{deg3aux} and~\eqref{ddo3} as
\begin{equation}
\o_3=\bigl\langle\tfrac13\bar\phi^3-(\bar W+\bar F)\bar\phi\bigr\rangle,\qquad
\dd\o_3=-\bigl\langle F^2\bigr\rangle.
\label{deg3}
\end{equation}
Up to a sign, $\dd\o_3$ is nothing but the second Chern character, and as such can be expressed as a derivative of a Chern-Simons (CS) 3-form,
\begin{equation}\label{defCS3}
\omega_3^\text{CS}=\bigl\langle FA-\tfrac13A^3\bigr\rangle,\qquad
\dd\omega_3^\text{CS}=\bigl\langle F^2\bigr\rangle.
\end{equation}
The form $\o_3+\o^\text{CS}_3$ is now closed and thus gives rise to a well-defined action in two spacetime dimensions~\cite{Hull:1990ms}. It is, however, not gauge invariant, for the CS form $\o^\text{CS}_3$ is not. Under an infinitesimal gauge transformation generated by $g=e^\epsilon$, the gauge field transforms as
\begin{equation}
\d A=-\dd\epsilon+[\epsilon,A],\qquad
\d F=[\epsilon,F],
\label{gaugetransfo}
\end{equation}
and the CS 3-form satisfies $\d\o^\text{CS}_3=-\langle\dd\epsilon\,\dd A\rangle$. Being itself a closed form, this is given by a derivative of a 2-form, which in turn determines the anomaly of the corresponding two-dimensional theory~\cite{Bertlmann:1996xk},
\begin{equation}
\d\o^\text{CS}_3=\dd\Aa_2,\qquad
\Aa_2=-\langle\epsilon\,\dd A\rangle.
\label{anomaly2}
\end{equation}

To find an explicit expression for the two-dimensional version of the WZ term, $\tilde\o_2$, is a more subtle problem. Namely, it turns out impossible to write $\tilde\o_2$ in the product basis built out of the MC 1-form with constant coefficients. This reflects the topological nature of the NG boson interactions, generated by the WZ term. What \emph{can} be given as a simple two-dimensional expression is the gauge-field-dependent part of the WZ term, defined by
\begin{equation}
\o_{3A}\equiv\o_3(A)-\o_3(0)+\o^\text{CS}_3.
\label{om3Adef}
\end{equation}
One can in other words construct a 2-form $\tilde\o_{2A}$ such that $\dd\tilde\o_{2A}=\o_{3A}$. Unlike the $\tilde\o_1$ of eq.~\eqref{CS1}, this is not left $G$-invariant though, and thus does not have a simple canonical form. An explicit manipulation leads to the result
\begin{equation}
\tilde\o_{2A}=-\bigl\langle(\bar A+\bar A_\parallel)\phi+\bar A_\parallel\bar A_\perp\bigr\rangle=-\bigl\langle(\bar A+\bar A_\parallel)\bar\phi-\bar A_\parallel\bar A_\perp\bigr\rangle.
\label{om2A}
\end{equation}
Here and in the following, we use the $\parallel$ and $\perp$ symbols to indicate the unbroken and broken components of elements of the Lie algebra of $G$, respectively.

The 3-forms~\eqref{deg3aux} and~\eqref{deg3} appeared previously in ref.~\cite{DHoker:1995it}. The two-dimensional matrix expression~\eqref{om2A} for the gauge-field-dependent part of the WZ term is, to the best of our knowledge, new. As pointed out in the introduction, the same 3-form $\o_3$ defines the anomalous Goldstone-Wilczek current in four spacetime dimensions~\cite{Goldstone:1981kk}.


\subsection{Degree 4 (three dimensions)}
\label{subsec:deg4}

Here the independent cohomology generators are parameterized by a matrix $m_{\a\b}$, which is required to be a constant symmetric tensor, invariant under the adjoint action of $H$, that is, it satisfies a constraint analogous to eq.~\eqref{Hinv}. By Schur's lemma, it is necessarily proportional to the unit matrix on every simple factor of $H$, and is thus uniquely determined by a single real coefficient on every such factor. On the Abelian part of $H$, on the other hand, $m_{\a\b}$ can take an arbitrary value. The associated closed gauge-invariant 4-form reads
\begin{equation}
\o_4=m_{\a\b}\bar W^\a\bar W^\b.
\label{deg4}
\end{equation}
Just like in the case of the degree-2 cohomology, this form can easily be written explicitly as a derivative of a 3-form, directly corresponding to a Lagrangian density in the three-dimensional physical spacetime,
\begin{equation}
\tilde\o_3=m_{\a\b}\bar\t^\a\bigl(\dd\bar\t^\b+\tfrac13f^\b_{\phantom\b\g\d}\bar\t^\g\bar\t^\d\bigr).
\label{CS3}
\end{equation}
This result appeared previously in ref.~\cite{DHoker:1995it} and was rederived using elementary field theory in ref.~\cite{Brauner:2014ata}. The Lagrangian~\eqref{CS3} is known to describe for instance the so-called Hopf invariant in two-dimensional magnets. It also generates a coupling of topologically nontrivial spin configurations (so-called skyrmions) to the electromagnetic field~\cite{Brauner:2014ata}.


\subsection{Degree 5 (four dimensions)}
\label{subsec:deg5}

In this case of most interest, the independent cohomology generators are parameterized by constant symmetric $G$-invariant tensors $d_{ABC}$ that vanish on the unbroken subgroup, that is, satisfy $d_{\a\b\g}=0$. The tensor $d_{ABC}$ is unique up to an overall scale for every simple compact Lie group. In fact, nonzero $d_{ABC}$ only exists for $\gr{SU}(N)$ with $N\geq3$, including $\gr{SO}(6)\simeq \gr{SU}(4)/\gr{Z}_2$. We can again use the compact trace notation~\eqref{tracenotation}, except that now, $d_{ABC}=\frac12\sum_jd_j\tr_j(T_A\{T_B,T_C\})$.\footnote{The coefficients $d_j$ can now of course differ from those appearing in eq.~\eqref{tracenotation}.} The corresponding gauge-invariant 5-form then reads
\begin{equation}
\o_5=\bigl\langle\tfrac1{10}\bar\phi^5-\tfrac12(\bar W+\bar F)\bar\phi^3+(\bar W^2+\bar F^2)\bar\phi+\tfrac12(\bar W\bar F+\bar F\bar W)\bar\phi\bigr\rangle.
\label{deg5}
\end{equation}
As in the case of the degree-3 cohomology, this form is actually not closed,
\begin{equation}
\dd\o_5=d_{ABC}F^AF^BF^C=\bigl\langle F^3\bigr\rangle,
\label{ddo5}
\end{equation}
which hints at a geometric obstruction to gauging the WZ term, related to an anomaly in the underlying microscopic theory. Thanks to $d_{\a\b\g}=0$, it is nevertheless possible to gauge the unbroken subgroup $H$ while maintaining the closedness of $\o_5$. Should gauge fields for the whole group $G$ be included, we can proceed as in the degree-3 case and note that the right-hand side of eq.~\eqref{ddo5} is the third Chern character, which is given by a derivative of a CS 5-form,
\begin{equation}\label{defCS5}
\o_5^\text{CS}=\bigl\langle F^2 A-\tfrac12FA^3+\tfrac1{10}A^5\bigr\rangle,\qquad
\dd\o_5^\text{CS}=\bigl\langle F^3\bigr\rangle.
\end{equation}
The form $\o_5-\o^\text{CS}_5$ is then closed and thus gives rise to a well-defined action in four spacetime dimensions. This four-dimensional theory suffers from an anomaly, though, which is given by an analogue of eq.~\eqref{anomaly2}
\begin{equation}
\d\o^\text{CS}_5=-\dd\Aa_4,\qquad
\Aa_4=\bigl\langle\epsilon\,\dd\bigl(A\dd A+\tfrac12A^3\bigr)\bigr\rangle.
\label{anomaly4}
\end{equation}
It is now possible to give a closed four-dimensional expression $\tilde\o_{4A}$ for the gauge-field-dependent part of the WZ term, defined similarly to eq.~\eqref{om3Adef} by\footnote{In ref.~\cite{DHoker:1995it}, a closed 5-form is constructed as $\omega_5(U,A)-\omega_5(\mathbb{1},A)$. This 5-form differs from $\omega_5(A)-\omega_5^\text{CS}$ by a derivative of a 4-form depending on the gauge fields only, implying a different representation for the anomaly.}
\begin{equation}
\o_{5A}\equiv\o_5(A)-\o_5(0)-\o^\text{CS}_5.
\end{equation}
and $\dd\tilde\o_{4A}=\o_{5A}$. The 5-form $\o_{5A}$ can be integrated using transgression methods~\cite{Hull:1990ms}. We give the result in two equivalent forms, written in terms of $\phi$ and $\bar\phi$, respectively,
\begin{align}
\notag
\tilde\o_{4A}=\bigl\langle&\tfrac12\phi^3(\bar A+{\bar A}_\parallel)+\tfrac14\phi{\bar A}_\perp\phi(\bar A+{\bar A}_\parallel)+\tfrac12\phi^2[{\bar A}_\perp,{\bar A}_\parallel]\\
\notag
&+\phi(\tfrac12{\bar A}_\perp^3+\tfrac34{\bar A}_\perp^2{\bar A}_\parallel+\tfrac34{\bar A}_\parallel{\bar A}_\perp^2+\tfrac12{\bar A}_\parallel^2{\bar A}_\perp+\tfrac12{\bar A}_\parallel{\bar A}_\perp{\bar A}_\parallel+\tfrac12{\bar A}_\perp{\bar A}_\parallel^2+{\bar A}_\parallel^3)\\
\notag
&+\tfrac12{\bar A}_\perp{\bar A}_\parallel^3-\tfrac12{\bar A}_\parallel{\bar A}_\perp^3-\tfrac14{\bar A}_\parallel{\bar A}_\perp{\bar A}_\parallel{\bar A}_\perp\\
\label{om4A}
&+\tfrac12\bar F[\bar A+\tfrac12{\bar A}_\parallel,\phi]+\tfrac12(\bar W+W)[\tfrac12\bar A+{\bar A}_\parallel,\phi]+(\tfrac12\bar F+\tfrac12\bar W+\tfrac14 W)[{\bar A}_\parallel,{\bar A}_\perp]\bigr\rangle\\
\notag
=\bigl\langle&\tfrac12\bar\phi^3(\bar A+{\bar A}_\parallel)-\tfrac14\bar\phi{\bar A}_\perp\bar\phi(\bar A+{\bar A}_\parallel)-\tfrac12\bar\phi^2[{\bar A}_\perp,{\bar A}_\parallel]\\
\notag
&+\bar\phi(\tfrac12{\bar A}_\perp^3+\tfrac34{\bar A}_\perp^2{\bar A}_\parallel+\tfrac34{\bar A}_\parallel{\bar A}_\perp^2+\tfrac12{\bar A}_\parallel^2{\bar A}_\perp+\tfrac12{\bar A}_\parallel{\bar A}_\perp{\bar A}_\parallel+\tfrac12{\bar A}_\perp{\bar A}_\parallel^2+{\bar A}_\parallel^3)\\
\notag
&-\tfrac12{\bar A}_\perp{\bar A}_\parallel^3+\tfrac12{\bar A}_\parallel{\bar A}_\perp^3+\tfrac14{\bar A}_\parallel{\bar A}_\perp{\bar A}_\parallel{\bar A}_\perp\\
\notag
&+\tfrac12\bar F[\bar A+\tfrac12{\bar A}_\parallel,\bar\phi]+\tfrac12(\bar W+W)[\tfrac12\bar A+{\bar A}_\parallel,\bar\phi]-(\tfrac14\bar F+\tfrac14\bar W+\tfrac12 W)[{\bar A}_\parallel,{\bar A}_\perp]\bigr\rangle.
\end{align}
The ungauged version of the 5-form~\eqref{deg5} appeared previously in ref.~\cite{DHoker:1995it}. The gauged version thereof, as well as the four-dimensional expression~\eqref{om4A} for the gauge-field-dependent part of the WZ term, is new, and so is its special case~\eqref{om4ASig} worked out below.


\subsection{Simplified expressions for symmetric coset spaces}
\label{subsec:symmetric}

The MC form $\bar\t$ is a universal building block for the construction of invariant actions. It is, however, not always the most convenient one due to the complicated transformation properties of the NG variable $U$, given by eq.~\eqref{coset}. A dramatic simplification can be achieved for symmetric coset spaces, that is, such $G$ and $H$ that admit an automorphism $\mathcal R$ of the Lie algebra under which $\mathcal{R}(T_\a)=T_\a$ and $\mathcal{R}(T_a)=-T_a$. We also need to choose a coset representative $U$ that is inverted by the automorphism $\mathcal R$; this can always be accomplished through the exponential parameterization, $U(x)=e^{\im\pi^a(x)T_a}$, in terms of the NG fields $\pi^a(x)$. It is then possible to define a field variable that transforms linearly under the whole group $G$~\cite{Coleman:1969sm,Callan:1969sn},
\begin{equation}
\Sigma(x)\equiv U(x)^2,\qquad
\Sigma\xrightarrow{g}g\Sigma\mathcal R(g)\inv.
\label{SigmaDef}
\end{equation}
The gauged MC form then effectively translates into a covariant derivative of $\Sigma$ through
\begin{equation}
\bar\phi=\tfrac12U\inv(D\Sigma)U\inv=-\tfrac12U(D\Sigma\inv)U,
\label{dictionary}
\end{equation}
where the covariant derivative of $\Sigma$ takes a very simple expression thanks to the linear transformation properties of $\Sigma$,
\begin{equation}\label{covSig}
D\Sigma\equiv\dd\Sigma+A\Sigma-\Sigma\RA,
\end{equation}
and we introduced a shorthand notation for the action of the automorphism $\R$, $\RA\equiv\R(A)$. An analogous relation to eq.~\eqref{dictionary} connecting $\phi$ and $\dd\Sigma$ holds. To complete the dictionary required to translate our results into the $\Sigma$-variable, one needs the relations
\begin{equation}
\bar W=\bar F_\parallel-\bar\phi^2,\qquad
W=-\phi^2,
\end{equation}
valid for symmetric coset spaces. The broken and unbroken components of various Lie-algebra-valued objects can be projected out using the automorphism $\mathcal R$.

In the following, we will focus on the cohomology generators of degree 3 and 5, which were demonstrated above to have a simple matrix form. The case of degree-2 and degree-4 generators is somewhat more complicated due to the mere $H$-invariance of the associated coefficients. We refer the reader to ref.~\cite{Brauner:2014ata} for a discussion of this case, including a matrix form of these generators under certain further assumptions on the symmetry algebra.

The 2-form $\tilde\o_{2A}$~\eqref{om2A} can now be rewritten as a simple polynomial in $\Sigma$ and the external gauge field,
\begin{equation}
\begin{split}
\tilde\o_{2A}&=\bigl\langle\tfrac34\dd\Sigma\Sigma\inv A-\tfrac14\dd\Sigma\inv\Sigma\RA-\tfrac12\Sigma\RA\Sigma\inv A\bigr\rangle\\
&=\bigl\langle\tfrac34D\Sigma\Sigma\inv A-\tfrac14D\Sigma\inv\Sigma\RA+\tfrac12\Sigma\RA\Sigma\inv A\bigr\rangle.
\end{split}
\label{om2ASig}
\end{equation}
Likewise, for the 4-form $\tilde\o_{4A}$~\eqref{om4A} we find
\begin{align}
\notag
\tilde\o_{4A}=\bigl\langle&-\tfrac{11}{32}\dd\Sigma\dd\Sigma\inv\dd\Sigma\Sigma\inv A+\tfrac{5}{32}\dd\Sigma\inv\dd\Sigma\dd\Sigma\inv\Sigma \RA+\tfrac{3}{32}\dd\Sigma \RA\dd\Sigma\inv A+\tfrac{11}{64}\dd\Sigma\Sigma\inv A\dd\Sigma\Sigma\inv A\\
\notag
&-\tfrac{5}{64}\dd\Sigma\inv\Sigma \RA\dd\Sigma\inv\Sigma \RA+\tfrac{1}{4}\dd\Sigma\dd\Sigma\inv\Sigma \RA\Sigma\inv A-\tfrac{1}{4}\dd\Sigma\inv\dd\Sigma\Sigma\inv A\Sigma \RA\\
\notag
&-\tfrac{9}{32}\dd\Sigma\Sigma\inv A\Sigma \RA\Sigma\inv A+\tfrac{7}{32}\dd\Sigma\inv\Sigma \RA\Sigma\inv A\Sigma \RA\\
\label{om4ASig}
&+\tfrac{3}{32}\dd\Sigma \RA\Sigma\inv A^2-\tfrac{5}{32}\dd\Sigma\inv A\Sigma \RA^2+\tfrac{5}{32}\dd\Sigma \RA^2\Sigma\inv A-\tfrac{3}{32}\dd\Sigma\inv A^2\Sigma \RA\\
\notag
&+\tfrac{13}{32}\dd\Sigma\Sigma\inv A^3-\tfrac{3}{32}\dd\Sigma\inv\Sigma \RA^3+\tfrac{1}{8}\Sigma \RA\Sigma\inv A\Sigma \RA\Sigma\inv A-\tfrac{5}{16}\Sigma \RA\Sigma\inv A^3+\tfrac{3}{16}\Sigma\inv A\Sigma \RA^3\\
\notag
&+\tfrac{1}{8}(-\dd\Sigma \RF\Sigma\inv A-\dd\Sigma \RA\Sigma\inv F+\dd\Sigma\inv F\Sigma \RA+\dd\Sigma\inv A\Sigma \RF)\\
\notag
&-\tfrac{7}{16}\dd\Sigma\Sigma\inv \{F,A\}+\tfrac{1}{16}\dd\Sigma\inv\Sigma \{\RF,\RA\}+\tfrac{3}{8}\Sigma \RA\Sigma\inv \{F,A\}-\tfrac{1}{8}\Sigma\inv A\Sigma \{\RF,\RA\}\bigr\rangle,
\end{align}
where $\RF\equiv \R(F)$. In principle, an equivalent result written in terms of covariant derivatives of $\Sigma$ can be obtained from the expression for $\tilde\o_{4A}$ in terms of $\bar\phi$. We choose not to spell out such a result here, as it is in no way simpler than eq.~\eqref{om4ASig}, and moreover obscures the dependence of $\tilde\o_{4A}$ on the gauge field.


\subsection{Adding a spectator $\gr{U(1)}$ symmetry}
\label{subsec:spectator}

The construction of WZ terms in ref.~\cite{DHoker:1995it}, on which our discussion is based, requires the symmetry group $G$ to be compact and semisimple. It can, however, easily be extended to situations where one adds a ``spectator'' $\gr{U}(1)_\text{s}$ symmetry, that is, considers symmetry-breaking patterns of the type $G\times\gr{U}(1)_\text{s}\to H\times\gr{U}(1)_\text{s}$, where $G$ itself is semisimple. This is of relevance for instance in QCD where the baryon number $\gr{U(1)}_\text{B}$ symmetry plays the role of such a spectator.

At first sight, it seems as if nothing changed by the presence of the unbroken spectator symmetry. The coset space keeps its original structure, $G/H$, and the corresponding NG fields are completely insensitive to the spectator symmetry. A gauge field $A_\text{s}$ for $\gr{U}(1)_\text{s}$ does not enter at all the cohomology generators on $G/H$, listed in section~\ref{sec:results}. However, it does give rise to an extra closed invariant 2-form,
\begin{equation}\label{omspec}
\o^\text{s}_2=F^\text{s}=\dd A^\text{s}.
\end{equation}
This can, in fact, be naturally incorporated in eq.~\eqref{deg2} given that $f^{\alpha}_{\phantom\alpha BC}=0$ holds for any generators $T_{B,C}$ when $\alpha$ corresponds to the $\gr{U}(1)_\text{s}$ generator. The 2-form \eqref{omspec} can then be used to construct new non-primitive cohomology generators on $[G\times\gr{U}(1)_\text{s}]/[H\times\gr{U}(1)_\text{s}]$,
\begin{equation}
\o_{d+1}^\text{s}=\o_2^\text{s}\wedge\o_{d-1}=F^\text{s}\wedge\o_{d-1},\qquad
\tilde\o_d^\text{s}=A^\text{s}\wedge\o_{d-1}.
\label{spectator}
\end{equation}
In the more familiar Lagrangian language, this amounts to a WZ term proportional to $A^\text{s}_\mu j^\mu$, where $j^\mu$ is a topological current generalizing the Goldstone-Wilczek current of ref.~\cite{Goldstone:1981kk}.


\section{Examples}
\label{sec:examples}


\subsection{Chiral coset spaces $(G_\text{L}\times G_\text{R})/G_\text{L+R}$}
\label{subsec:complex}

With the special choice $G=\gr{SU}(N)$, this coset space plays a prominent role in QCD, or generally in QCD-like theories with quarks transforming in a complex representation of the gauge group. However, the results spelled out below hold equally well for any compact simple $G$ satisfying the assumptions on the topology of the coset space. This coset space is symmetric thanks to the automorphism that swaps the corresponding generators of $G_\text{L}$ and $G_\text{R}$. The conditions $d_{\a\b}=0$ and $d_{\a\b\g}=0$ require that, up to an overall rescaling, the $\langle\cdot\rangle$ operation~\eqref{tracenotation} is defined as trace over the Lie algebra of $G_\text{L}$ minus trace over that of $G_\text{R}$.

We introduce the notation $(T_{\text{L},A},T_{\text{R},B})$ for the generators of the $G_\text{L}\times G_\text{R}$ symmetry group, where both entries in the parentheses now run over the generators of $G$. With a slight abuse of notation, the linearly transforming matrix variable~\eqref{SigmaDef} can be denoted as $(\Sigma,\Sigma\inv)$, where $\Sigma\in G$.\footnote{This notation is motivated by the conventions commonly used in the chiral perturbation theory of QCD.} The gauge field for the full chiral group then corresponds to pairs $(A_\text{L},A_\text{R})$, where both $A_\text{L}$ and $A_\text{R}$ take values in the Lie algebra of $G$. With this notation, eq.~\eqref{om2ASig} gives
\begin{equation}
\begin{split}
\tilde\o_{2A}&=\tr\bigl(\dd\Sigma\Sigma\inv A_\text{L}-\dd\Sigma\inv\Sigma A_\text{R}-\Sigma A_\text{R}\Sigma\inv A_\text{L}\bigr)\\
&=\tr\bigl(D\Sigma\Sigma\inv A_\text{L}-D\Sigma\inv\Sigma A_\text{R}+\Sigma A_\text{R}\Sigma\inv A_\text{L}\bigr),
\end{split}
\label{om2ALR}
\end{equation}
where $D\Sigma\equiv\dd\Sigma+A_\text{L}\Sigma-\Sigma A_\text{R}$ is implied by eq.~\eqref{covSig}. Likewise, eq.~\eqref{om4ASig} leads to
\begin{align}
\notag
\tilde\o_{4A}=\tr\bigl(&-\tfrac12\dd\Sigma\dd\Sigma\inv\dd\Sigma\Sigma\inv A_\text{L}+\tfrac14\dd\Sigma\Sigma\inv A_\text{L}\dd\Sigma\Sigma\inv A_\text{L}+\tfrac12\dd\Sigma\dd\Sigma\inv\Sigma A_\text{R} \Sigma\inv A_\text{L}\\
\label{om4ALR}
&-\tfrac12\dd\Sigma\Sigma\inv A_\text{L}\Sigma A_\text{R}\Sigma\inv A_\text{L}+\tfrac14\dd\Sigma A_\text{R}\Sigma\inv A_\text{L}^2+\tfrac14\dd\Sigma A_\text{R}^2\Sigma\inv A_\text{L}+\tfrac12\dd\Sigma\Sigma\inv A_\text{L}^3\\
\notag
&+\tfrac18\Sigma A_\text{R}\Sigma\inv A_\text{L}\Sigma A_\text{R}\Sigma\inv A_\text{L}-\tfrac12\Sigma A_\text{R}\Sigma\inv A_\text{L}^3-\tfrac14\dd\Sigma F_\text{R}\Sigma\inv A_\text{L}-\tfrac14\dd\Sigma A_\text{R}\Sigma\inv F_\text{L}\\
\notag
&-\tfrac12\dd\Sigma\Sigma\inv\{F_\text{L},A_\text{L}\}+\tfrac12\Sigma A_\text{R}\Sigma\inv\{F_\text{L},A_\text{L}\}\bigr)-(\Sigma\leftrightarrow\Sigma\inv,\,\text{L}\leftrightarrow\text{R}).
\end{align}
This expression agrees with classic results available in the literature, taking into account that the 4-form may be varied by adding a derivative of a 3-form without changing the action: $\tilde\o_{4A}+\frac14\dd\bigl[\tr(\Sigma A_\text{R}\Sigma\inv\dd A_\text{L}-\Sigma\inv A_\text{L}\Sigma\dd A_\text{R})\bigr]$ agrees, up to an overall normalization factor, with refs.~\cite{AlvarezGaume:1984dr,Manohar:1984uq}, whereas $\tilde\o_{4A}-\frac14\dd\bigl[\tr(\dd\Sigma A_\text{R}\Sigma\inv A_\text{L}-\dd\Sigma\inv A_\text{L}\Sigma A_\text{R})\bigr]$ agrees, again up to overall normalization, with refs.~\cite{Kaymakcalan:1983qq,Manes:1984gk}.

Finally, let us comment on our assumption on the triviality of $\pi_d(G/H)$ for the current case of the $(G_\text{L}\times G_\text{R})/G_\text{L+R}$ coset space. Since this coset space has the same topology as $G$ itself, the situation is very simple for $d=2$ where $\pi_2(G)=0$ for any compact connected Lie group~\cite{Weinberg:1996v2}. No such generic claim holds for $\pi_4(G)$ though. On the other hand, as remarked above, a nonzero symmetric invariant tensor $d_{ABC}$, and thus a nontrivial primitive cohomology generator $\omega_5$, only exists for $G=\gr{SU}(N)$ with $N\geq3$, for which $\pi_4(G)$ is trivial~\cite{Weinberg:1996v2}. Let us add for completeness that $\pi_4(G)=0$ also for $G=\gr{SO}(N)$ with $N\geq 6$ and for the exceptional groups $\gr{E}_{6,7,8}$, $\gr{F}_4$, $\gr{G}_2$; for all the other compact simple Lie groups, $\pi_4(G)$ is nontrivial~\cite{Weinberg:1996v2,Abanov:2017zok}.


\subsubsection{Application: QCD in external electromagnetic field}

The behavior of quark matter under strong magnetic fields has been subject to intensive investigations~\cite{Andersen:2014xxa,Miransky:2015ava}. A number of fascinating phenomena that occur in such a medium owe their existence to the chiral anomaly. For a concrete illustration of the construction of WZ terms, we shall therefore look at QCD coupled to gauge fields for two unbroken, mutually commuting conserved charges, namely the electric charge $Q$ and the baryon number $B$. The corresponding gauge potentials will be denoted as $A$ and $A^B$ in accord with the notation introduced in ref.~\cite{Son:2007ny}.

Let us start with the case of two light quark flavors, where the full chiral symmetry group is $\gr{SU}(2)_\text{L}\times\gr{SU}(2)_\text{R}$ and its unbroken subgroup $\gr{SU}(2)_\text{L+R}$. Since the fully symmetric symbol $d_{ABC}$ vanishes for $\gr{SU(2)}$, the primitive WZ term $\o_5$ is now absent. It is, however, possible to construct a degree-5 cohomology generator using the spectator baryon number symmetry $\gr{U(1)}_\text{B}$, as explained in section~\ref{subsec:spectator}. We start by noting that the operator of electric charge in the quark flavor space, $Q=\text{diag}(\tfrac23,-\tfrac13)$, is \emph{not} a generator of the unbroken subgroup $\gr{SU}(2)_\text{L+R}$. It can nevertheless be related to one of its generators, $I_3$ (isospin), through the Gell-Mann-Nishijima relation $Q=I_3+\frac B2$. The total gauge potential of the system can then be cast as
\begin{equation}
A^BB+AQ=\left(A^B+\frac A2\right)B+AI_3.
\end{equation}
This indicates that there is a non-primitive degree-5 cohomology generator, corresponding to a WZ term in four spacetime dimensions, given via eq.~\eqref{spectator} by
\begin{equation}
\tilde\o_4=\left(A^B+\frac A2\right)\wedge\o_3,
\end{equation}
where $\o_3$ is constructed using the matrix gauge field $AI_3$. To carry out this construction, we need an expression for $\o_3$ valid for chiral coset spaces, which follows from eq.~\eqref{deg3},
\begin{equation}
\o_3=\tr\bigl(-\tfrac13D\Sigma D\Sigma\inv D\Sigma\Sigma\inv-D\Sigma\Sigma\inv F_\text{L}+D\Sigma\inv\Sigma F_\text{R}\bigr).
\label{deg3Sigma}
\end{equation}
Upon translating this from the language of differential forms and restricting the gauge field to $AI_3$, this leads to the gauge-invariant expression for the Goldstone-Wilczek current,
\begin{equation}
j^\m\propto\epsilon^{\m\n\a\b}\bigl\{\tr\bigl[(\Sigma D_\n\Sigma\inv)(\Sigma D_\a\Sigma\inv)(\Sigma D_\b\Sigma\inv)\bigr]-\tfrac32F_{\n\a}\tr\bigl[I_3(\Sigma D_\b\Sigma\inv+D_\b\Sigma\inv\Sigma)\bigr]\bigr\},
\label{GWcurrent}
\end{equation}
which agrees with ref.~\cite{Son:2007ny} up to an overall normalization, not fixed here.\footnote{To compare our results, one has multiply the gauge fields in ref.~\cite{Son:2007ny} by $\im$ and set $e=-1$ therein.}

Let us now move on to the case of three light quark flavors. In this case, both the primitive WZ term $\o_5$ and the non-primitive one $\o_2^\text{s}\wedge\o_3$ exist. To get the primitive WZ term, we employ eq.~\eqref{om4ALR}. Since a single generator of the chiral group $\gr{SU}(3)_\text{L}\times\gr{SU}(3)_\text{R}$, corresponding to the electric charge operator, is gauged, all terms with more than one factor of $A$ therein vanish identically. With the shorthand notation
\begin{equation}
L\equiv\Sigma\dd\Sigma\inv,\qquad
R\equiv\dd\Sigma\inv\Sigma,
\end{equation}
the form $\tilde\o_{4A}$ then simplifies to
\begin{equation}
\tilde\o_{4A}\to\tr\bigl(-\tfrac12L^3A+\tfrac14L\Sigma F\Sigma\inv A+\tfrac14L\Sigma A\Sigma\inv F+\tfrac12L\{F,A\}\bigr)-(\Sigma\leftrightarrow\Sigma\inv,L\leftrightarrow-R).
\end{equation}
This corresponds to the Lagrangian density
\begin{equation}
\begin{split}
\La\propto\epsilon^{\m\n\a\b}\bigl[&A_\m\tr(QL_\n L_\a L_\b+QR_\n R_\a R_\b)-F_{\m\n}A_\a\tr(Q^2L_\b+Q^2R_\b)\\
&-\tfrac12F_{\m\n}A_\a\tr(Q\Sigma Q\partial_\b\Sigma\inv-Q\Sigma\inv Q\partial_\b\Sigma)\bigr],
\end{split}
\end{equation}
where now $Q=\text{diag}(\frac23,-\frac13,-\frac13)$. The non-primitive generator again follows from eq.~\eqref{deg3Sigma}. The expression for the Goldstone-Wilczek current~\eqref{GWcurrent} is modified owing to the fact that the electric charge operator $Q$ now does belong to the Lie algebra of the unbroken vector subgroup,
\begin{equation}
j^\m\propto\epsilon^{\m\n\a\b}\bigl\{\tr\bigl[(\Sigma D_\n\Sigma\inv)(\Sigma D_\a\Sigma\inv)(\Sigma D_\b\Sigma\inv)\bigr]-\tfrac32F_{\n\a}\tr\bigl[Q(\Sigma D_\b\Sigma\inv+D_\b\Sigma\inv\Sigma)\bigr]\bigr\}.
\end{equation}
The corresponding contribution to the anomalous Lagrangian density is simply $A^B_\m j^\m$. Both of the above results for QCD with three quark flavors agree with ref.~\cite{Son:2007ny}.


\subsection{$\gr{SU}(2N)/\gr{SO}(2N)$ and $\gr{SU}(2N)/\gr{Sp}(2N)$ coset spaces}
\label{subsec:pseudoreal}

These coset spaces appear for instance in the study of QCD-like theories with quarks in a real or pseudoreal representation of the gauge group, both in the context of strong interaction physics~\cite{Kogut:1999iv,Kogut:2000ek} and in models of dynamical electroweak symmetry breaking~\cite{Peskin:1980gc}. (Pseudo)rea\-li\-ty guarantees that the theory of $N$ flavors of massless quarks possesses an enhanced $\gr{SU}(2N)$ flavor symmetry. In case of a real representation, this is spontaneously broken in the vacuum to the $\gr{SO}(2N)$ subgroup, whereas in the pseudoreal case it is broken down to the $\gr{Sp}(2N)$ subgroup. The underlying conditions on the topology of the coset space are easy to check at least in the pseudoreal case with $N=2$: $\gr{SU}(4)/\gr{Sp}(4)\simeq \gr{SO}(6)/\gr{SO}(5)\simeq S^5$, and hence both $\pi_2(G/H)$ and $\pi_4(G/H)$ are trivial in this case. For other coset spaces in this class, the homotopy condition may not be satisfied though.\footnote{Ref.~\cite{Davighi:2018xwn} analyses a variant of the $\gr{SO}(6)/\gr{SO}(4)$ coset space, for which several new topological terms, not covered by the homotopy classification of~\cite{DHoker:1995it}, are reported. Let us, however, note that our coset space $\gr{SU(4)}/\gr{SO(4)}\simeq\gr{SO(6)}/\gr{SO(4)}$ differs from that studied in ref.~\cite{Davighi:2018xwn} by the way the unbroken subgroup $H$ is embedded in the full group $G$. Indeed, our coset space, unlike that of ref.~\cite{Davighi:2018xwn}, is symmetric, and the topology properties may accordingly differ.}

Both coset spaces can be defined by the following relations, distinguishing unbroken and broken generators,
\begin{equation}
T_\a^T\Sigma_0+\Sigma_0T_\a=0,\qquad
T_a^T\Sigma_0-\Sigma_0T_a=0,
\label{pseudoantisymmetry}
\end{equation}
where $\Sigma_0$ is a fixed real unitary matrix, characterizing the ground state of the system. In the real case, $\Sigma_0$ is symmetric, whereas in the pseudoreal case, it is antisymmetric. It can be chosen for instance in the block form 
\begin{equation}
\Sigma_0=\begin{pmatrix}0&\mathbb{1}\\\mathbb{1}&0\end{pmatrix}\qquad\text{(real)},\qquad\qquad
\Sigma_0=\begin{pmatrix}0&\mathbb{1}\\-\mathbb{1}&0\end{pmatrix}\qquad\text{(pseudoreal)}.
\end{equation}
Both coset spaces are symmetric and the corresponding automorphism takes the form
\begin{equation}
\mathcal R(T_A)=-\Sigma_0T_A^T\Sigma_0\inv.
\end{equation}

Recalling that $G=\gr{SU}(2N)$ is simple, there is a unique $G$-invariant tensor $d_{AB}$, given up to an overall factor by $\tr(T_AT_B)$. This, however, does not vanish on the unbroken subgroup. Hence for the $\gr{SU}(2N)/\gr{SO}(2N)$ and $\gr{SU}(2N)/\gr{Sp}(2N)$ coset spaces, there is no nontrivial cohomology generator $\o_3$ of degree 3.

Thanks to $G$ being simple and unitary, there is likewise a unique $G$-invariant tensor $d_{ABC}$ for any $N\geq2$, corresponding in the angular bracket notation~\eqref{tracenotation} to a simple trace. The condition $d_{\a\b\g}=0$ is now satisfied automatically thanks to the (pseudo)antisymmetry property~\eqref{pseudoantisymmetry} of the unbroken generators, hence a single cohomology generator of degree 5 exists. Its physical, four-dimensional form $\tilde\o_{4A}$ is given by eq.~\eqref{om4ASig}, which can be considerably simplified by using the conjugation property of the linearly transforming $2N\times2N$ matrix variable $\Sigma=U^2$,
\begin{equation}
\Sigma^T=\Sigma_0\Sigma\Sigma_0\inv.
\end{equation}
We thus obtain
\begin{align}
\notag
\tilde\o_{4A}=\tr\bigl(&-\tfrac12\dd\Sigma\dd\Sigma\inv\dd\Sigma\Sigma\inv A+\tfrac14\dd\Sigma\Sigma\inv A\dd\Sigma\Sigma\inv A+\tfrac14\dd\Sigma\dd\Sigma\inv\Sigma \RA\Sigma\inv A-\tfrac14\dd\Sigma\inv\dd\Sigma\Sigma\inv A\Sigma\RA\\
\label{om4Areal}
&-\tfrac12\dd\Sigma\Sigma\inv A\Sigma \RA\Sigma\inv A-\tfrac12\dd\Sigma\Sigma\inv A^3+\tfrac18\Sigma \RA\Sigma\inv A\Sigma \RA\Sigma\inv A+\tfrac12\Sigma \RA\Sigma\inv A^3\\
\notag
&-\tfrac14\dd\Sigma \RA\Sigma\inv\dd A+\tfrac14\dd\Sigma\inv A\Sigma\dd\RA-\tfrac12\dd\Sigma\Sigma\inv\{\dd A,A\}+\tfrac12\Sigma \RA\Sigma\inv\{\dd A,A\}\bigr).
\end{align}
This agrees with the result for the $\gr{SU}(2N)/\gr{Sp}(2N)$ coset space, previously published in refs.~\cite{Duan:2000dy,Molinaro:2017mwb}, except for the sign of the $\Sigma \RA\Sigma\inv A\Sigma \RA\Sigma\inv A$ term. We also note that the contributions to the WZ term $\tilde\o_{4A}$ linear in the gauge field for a $\gr{U(1)}$ subgroup of $\gr{SU}(2N)$ were discussed, for both coset spaces, in ref.~\cite{Hochberg:2015vrg}.

Note that in the literature, it is more common to work with the variable
\begin{equation}
\tilde\Sigma(x)\equiv\Sigma(x)\Sigma_0,\qquad
\tilde\Sigma\xrightarrow{g}g\tilde\Sigma g^T.
\end{equation}
This is unitary and (anti)symmetric in the case of (pseudo)real fermions. It is straightforward to rewrite eq.~\eqref{om4Areal} in terms of $\tilde\Sigma$. We thus get
\begin{align}
\notag
\tilde\o_{4A}=\tr\bigl(&-\tfrac12\dd\tilde\Sigma\dd\tilde\Sigma\inv\dd\tilde\Sigma\tilde\Sigma\inv A+\tfrac14\dd\tilde\Sigma\tilde\Sigma\inv A\dd\tilde\Sigma\tilde\Sigma\inv A-\tfrac14\dd\tilde\Sigma\dd\tilde\Sigma\inv\tilde\Sigma A^T\tilde\Sigma\inv A+\tfrac14\dd\tilde\Sigma\inv\dd\tilde\Sigma\tilde\Sigma\inv A\tilde\Sigma A^T\\
&+\tfrac12\dd\tilde\Sigma\tilde\Sigma\inv A\tilde\Sigma A^T\tilde\Sigma\inv A-\tfrac12\dd\tilde\Sigma\tilde\Sigma\inv A^3+\tfrac18\tilde\Sigma A^T\tilde\Sigma\inv A\tilde\Sigma A^T\tilde\Sigma\inv A-\tfrac12\tilde\Sigma A^T\tilde\Sigma\inv A^3\\
\notag
&+\tfrac14\dd\tilde\Sigma A^T\tilde\Sigma\inv\dd A-\tfrac14\dd\tilde\Sigma\inv A\tilde\Sigma\dd A^T-\tfrac12\dd\tilde\Sigma\tilde\Sigma\inv\{\dd A,A\}-\tfrac12\tilde\Sigma A^T\tilde\Sigma\inv\{\dd A,A\}\bigr).
\end{align}


\subsection{$\gr{U}(N)/\gr{U}(N-1)$ coset spaces}

This class of coset spaces can be thought of as a generalization of the Higgs sector of the standard model. It is most easily visualized as the vacuum manifold, arising from breaking the $G=\gr{U}(N)$ symmetry down to $H=\gr{U}(N-1)$ by a nonzero expectation value of a complex scalar field $\Phi$, transforming in the fundamental representation of the $\gr{SU}(N)$ subgroup of $G$. This makes it obvious that the coset space is topologically equivalent to a sphere, $S^{2N-1}$.

Except for the trivial case $N=1$, which will be omitted here, these coset spaces are \emph{not} symmetric. We will tacitly ignore the fact that they do not satisfy our assumption on $G$ being semisimple, and thus the classification of WZ terms outlined in section~\ref{sec:results} is not necessarily exhaustive for them. It is nevertheless a nontrivial illustration of our general results that they can reproduce existing expressions for WZ terms in case of $N=2,3$~\cite{Hill:2007nz,Hill:2009wp}.

We start by inspecting whether or not nontrivial $G$-invariant symmetric tensors $d_{AB}$ and $d_{ABC}$, vanishing on the unbroken subgroup, exist. It is easy to see that $G$-invariance forces $d_{AB}$ to be a linear combination of $\tr(T_AT_B)$ and $\tr T_A\tr T_B$. This cannot, however, vanish on $H$ except for the $N=2$ case, for which there is a unique solution up to an overall factor,
\begin{equation}
d_{AB}=\tr(T_AT_B)-\tr T_A\tr T_B.
\label{dAB}
\end{equation}
As to $d_{ABC}$, it is likewise easy to see that $G$-invariance forces it to be a linear combination of $\tr(T_A,\{T_B,T_C\})$, $\tr T_A\tr T_B\tr T_C$, and $\tr T_A\tr(T_BT_C)$ summed over cyclic permutations of the indices. For $N\geq4$, the condition on vanishing of $d_{\a\b\g}$ has no nontrivial solution. For $N=3$, there is a unique solution up to an overall factor,
\begin{equation}
d_{ABC}=\tr(T_A\{T_B,T_C\})-[\tr T_A\tr(T_BT_C)+\tr T_B\tr(T_CT_A)+\tr T_C\tr(T_AT_B)]+\tr T_A\tr T_B\tr T_C.
\label{dABC}
\end{equation}
A nontrivial solution for $d_{ABC}$ also exists for $N=2$. In this case, however, the coset space $G/H$ is three-dimensional, hence there is obviously no primitive cohomology generator of degree 5.

We conclude that we can only expect nontrivial WZ terms for $N=2,3$. For $N=2$, we will have a unique cohomology generator of degree 3, whereas for $N=3$, we will have a unique cohomology generator of degree 5.\footnote{This is consistent with the fact that the coset space $\gr{U}(N)/\gr{U}(N-1)$ is topologically equivalent to $S^{2N-1}$, and that any $n$-sphere $S^n$ is known to only possess nontrivial de Rham cohomology generators of degrees $0$ and $n$. The latter is unique and corresponds to the volume form on the sphere.} Note that the topology condition on the coset space is satisfied in both cases: for $N=2$, we have $\pi_2(G/H)=\pi_2(S^3)=0$, and for $N=3$, we likewise have $\pi_4(G/H)=\pi_4(S^5)=0$.

Let us now see how to construct the degree-3 cohomology generator $\o_3$, or rather its two-dimensional form $\tilde\o_{2A}$, for $N=2$. The result is most conveniently expressed in terms of the linearly transforming complex scalar doublet $\Phi$ normalized so that $\he\Phi\Phi=1$. To that end, we introduce the vacuum expectation value $\varphi\equiv\langle0|\Phi|0\rangle$, and take advantage of the projector $\P\equiv\varphi\he\varphi$ to project out the unbroken and broken components of the MC form,
\begin{equation}
\bar V=(1-\P)\bar\t(1-\P),\qquad
\bar\phi=\bar\t-\bar V=\bar\t\P+\P\bar\t-\P\bar\t\P.
\end{equation}
We can now use all our general results for the WZ forms, expressed in terms of the MC form, and at the end of the day identify $\Phi=U\varphi$. Equation~\eqref{om2A} thus gives
\begin{equation}
\tilde\o_{2A}=-\he\Phi A\dd\Phi-\dd\he\Phi A\Phi+2A_0\he\Phi\dd\Phi+A_0\he\Phi A\Phi,
\end{equation}
where $A_0\equiv\tr A$. 
Our result for $\tilde\o_{2A}$ matches that given in ref.~\cite{Hill:2009wp} upon taking into account the differences in the used notation: in ref.~\cite{Hill:2009wp}, the gauge fields carry an extra factor of $-\im$, $A$ refers to the traceless $\gr{SU}(2)$ part of the gauge field rather than the full $\gr{U}(2)$ gauge field, and finally our definitions of $A_0$ differ by a factor of 2.

The WZ form $\tilde\o_{4A}$ in case of $N=3$ can be obtained in the same way. Its evaluation is however somewhat tedious, and it turns out to be more practical to start from the gauged 5-form~\eqref{deg5} and obtain $\tilde\o_{4A}$ by its integration, rather than to use eq.~\eqref{om4A} directly. This underlines the utility of both eq.~\eqref{deg5} and eq.~\eqref{om4A} for practical applications. Using eqs.~\eqref{deg5} and~\eqref{dABC} gives
\begin{equation}
\begin{split}
\o_5={}&2\he\Phi D\Phi D\he\Phi D\Phi D\he\Phi D\Phi-2F_0\he\Phi D\Phi D\he\Phi D\Phi+2D\he\Phi FD\Phi\he\Phi D\Phi\\
&+\he\Phi FD\Phi D\he\Phi D\Phi-D\he\Phi F\Phi D\he\Phi D\Phi+F_0^2\he\Phi D\Phi-(\tr F^2)\he\Phi D\Phi\\
&-F_0\he\Phi FD\Phi+F_0D\he\Phi F\Phi+\he\Phi F^2D\Phi-D\he\Phi F^2\Phi,
\end{split}
\end{equation}
where
\begin{equation}
F_0\equiv\tr F,\qquad
D\Phi\equiv\dd\Phi+A\Phi,\qquad
D\he\Phi\equiv\dd\he\Phi-\he\Phi A.
\end{equation}
The ensuing result for $\tilde\o_{4A}$ is best organized as a polynomial in the gauge field,
\begin{equation}
\tilde\o_{4A}=\tilde\o_{4A}^{(1)}+\tilde\o_{4A}^{(2)}+\tilde\o_{4A}^{(3)}+\tilde\o_{4A}^{(4)}.
\end{equation}
The individual pieces read
\begin{align}
\notag
\tilde\o_{4A}^{(1)}={}&-2A_0\he\Phi\dd\Phi\dd\he\Phi\dd\Phi+\he\Phi A\dd\Phi\dd\he\Phi\dd\Phi+\dd\he\Phi A\Phi\dd\he\Phi\dd\Phi-2\dd\he\Phi A\dd\Phi\he\Phi\dd\Phi,\\
\notag
\tilde\o_{4A}^{(2)}={}&A_0\dd A_0\he\Phi\dd\Phi-\tr(A\dd A)\he\Phi\dd\Phi-\tfrac12A_0\he\Phi\dd A\dd\Phi+\tfrac12A_0\dd\he\Phi\dd A\Phi\\
\notag
&-\tfrac12\dd A_0\he\Phi A\dd\Phi-\tfrac12\dd A_0\dd\he\Phi A\Phi+\tfrac12\he\Phi\{\dd A,A\}\dd\Phi+\tfrac12\dd\he\Phi\{\dd A,A\}\Phi\\
\notag
&-A_0\he\Phi A\Phi\dd\he\Phi\dd\Phi+A_0\he\Phi A\dd\Phi\he\Phi\dd\Phi-A_0\dd\he\Phi A\Phi\he\Phi\dd\Phi-\dd\he\Phi A\dd\Phi\he\Phi A\Phi\\
&+\he\Phi A^2\dd\Phi\he\Phi\dd\Phi-\dd\he\Phi A^2\Phi\he\Phi\dd\Phi-\tfrac12(\he\Phi A\dd\Phi)^2+\tfrac12(\dd\he\Phi A\Phi)^2,\\
\notag
\tilde\o_{4A}^{(3)}={}&\tfrac23A_0\dd A_0\he\Phi A\Phi-\tfrac23\tr(A\dd A)\he\Phi A\Phi-\tfrac13A_0\he\Phi\{\dd A,A\}\Phi-\tfrac13A_0\he\Phi A^2\dd\Phi+\tfrac13A_0\dd\he\Phi A^2\Phi\\
\notag
&+\tfrac23A_0\he\Phi A^2\Phi\he\Phi\dd\Phi+\tfrac23A_0\he\Phi A\dd\Phi\he\Phi A\Phi-\tfrac23A_0\dd\he\Phi A\Phi\he\Phi A\Phi-\tfrac23(\tr A^3)\he\Phi\dd\Phi\\
\notag
&+\tfrac23\he\Phi A^3\dd\Phi+\tfrac23\dd\he\Phi A^3\Phi+\tfrac13\he\Phi\dd A A^2\Phi-\tfrac13\he\Phi A^2\dd A\Phi+\tfrac23\he\Phi A^3\Phi\he\Phi\dd\Phi\\
\notag
&+\tfrac23\he\Phi A^2\dd\Phi\he\Phi A\Phi-\tfrac23\dd\he\Phi A^2\Phi\he\Phi A\Phi-\tfrac13\he\Phi A\dd\Phi\he\Phi A^2\Phi-\tfrac13\dd\he\Phi A\Phi\he\Phi A^2\Phi,\\
\notag
\tilde\o_{4A}^{(4)}={}&-\tfrac12A_0\he\Phi A^3\Phi+\tfrac12A_0\he\Phi A^2\Phi\he\Phi A\Phi-\tfrac12(\tr A^3)\he\Phi A\Phi+\tfrac12\he\Phi A^3\Phi\he\Phi A\Phi.
\end{align}
To compare this to the result published in ref.~\cite{Hill:2009wp}, recall that the WZ 4-form $\tilde\o_{4A}$ is only defined up to addition of a gauge-invariant 4-form and of a derivative of a 3-form. It is straightforward to show that
\begin{equation}
\begin{split}
\tilde\o_{4A}&-\he\Phi FD\Phi\he\Phi D\Phi+D\he\Phi F\Phi\he\Phi D\Phi\\
&+\dd\bigl[\tfrac12\he\Phi A\Phi(\he\Phi A\dd\Phi+\dd\he\Phi A\Phi)+\he\Phi A\dd\Phi\he\Phi\dd\Phi+\dd\he\Phi A\Phi\he\Phi\dd\Phi-\tfrac13A_0\he\Phi A\Phi\he\Phi\dd\Phi\bigr]
\end{split}
\end{equation}
agrees with eq.~(58) of ref.~\cite{Hill:2009wp} up to an overall factor, taking into account the differences in the notation used.


\subsection{Group manifolds $G/\{\}$}

Since existing results for coset spaces that are not symmetric are sparse, we conclude our list of examples with a class of WZ terms that, to the best of our knowledge, have not been pointed out before. Consider the coset spaces, for which a non-Abelian symmetry group $G$ is fully spontaneously broken. The case of $G=\gr{SU(2)}$ is relevant for instance for the so-called canted (anti)ferromagnets~\cite{Roman:canted}. We will, however, keep the discussion general, only taking into account the restriction on the homotopy groups required by our construction of the WZ terms. The discussion of the topology constraints is similar to the case of chiral coset spaces (see section~\ref{subsec:complex}): $\pi_2(G)=0$ for any compact connected Lie group and $\pi_4(G)=0$ for $\gr{SU}(N)$ with $N\geq 3$, which is the only case where a nonzero symmetric invariant tensor $d_{ABC}$ exists. Note that thanks to the triviality of $H$, the vanishing of $d_{\a\b}$ and $d_{\a\b\g}$ does not impose any further constraints.

When $H=\{\}$, we can replace $\phi$ with $\t$ and $\bar A_\perp$ with $\bar A$, and drop all terms containing $W$, $\bar W$ or $\bar A_\parallel$. The final result is most easily expressed in terms of the variable
\begin{equation}
\tilde\t\equiv U\t U\inv=\dd UU\inv.
\end{equation}
For the 2-form $\tilde\o_{2A}$ and the 4-form $\tilde\o_{4A}$ we thus get respectively from eqs.~\eqref{om2A} and~\eqref{om4A},
\begin{equation}
\begin{split}
\tilde\o_{2A}&=\bigl\langle\tilde\t A\bigr\rangle,\\
\tilde\o_{4A}&=\bigl\langle\tfrac12\tilde\t^3A+\tfrac14\tilde\t A\tilde\t A-\tfrac12\tilde\t A^3-\tfrac12\tilde\t\{\dd A,A\}\bigr\rangle.
\end{split}
\end{equation}


\section{Summary and discussion}
\label{sec:discussion}


In this paper, we have worked out explicit expressions for gauged WZ terms for a general coset space $G/H$ satisfying a moderate topology assumption, based on the construction of ref.~\cite{DHoker:1995it}. The main novel result obtained here is the expression~\eqref{om4A} for the gauge-field-dependent part of the WZ term in four spacetime dimensions, and its special case~\eqref{om4ASig} valid for symmetric coset spaces. Aiming at practitioners working on diverse physical applications, we restricted the amount of formalism to a minimum and merely listed the results, including several concrete examples for specific choices of the symmetry group $G$ and its unbroken subgroup $H$. Some concluding remarks are in order here.


\subsection{General structure of WZ terms}

As the results listed in section~\ref{sec:results} show, there is an important conceptual difference between the even- and odd-degree WZ terms. The even-degree ones, given by eqs.~\eqref{deg2} and \eqref{deg4}, can be completely gauged. However, they are determined by $H$-invariant tensor coefficients, which means that they do not have a simple matrix form in terms of generators of the whole group $G$. In addition, the one- and three-dimensional forms of these WZ terms, $\tilde\o_1$~\eqref{CS1} and $\tilde\o_3$~\eqref{CS3}, are given solely in terms of $\bar\t^\a$. This is not a coincidence. Namely, when the WZ term can be fully gauged, it can be mapped on a Chern-Simons-type action for the $H$-valued gauge field $\bar V=\bar\t^\a T_\a$~\cite{Brauner:2014ata}. Indeed, the 3-form in eq.~\eqref{CS3} corresponds to the non-Abelian Chern-Simons theory for $\bar\t^\a$.

The odd-degree WZ terms~\eqref{deg3} and~\eqref{deg5}, on the contrary, do have a simple matrix form owing to the fact that they are determined by $G$-invariant tensor coefficients. However, they cannot be completely gauged while preserving full $G$-invariance, which reflects their physical origin in the chiral anomaly. There is no simple expression for the corresponding two- and four-dimensional forms $\tilde\o_2$ and $\tilde\o_4$, respectively.


\subsection{Ambiguity of the WZ terms}

The WZ forms $\tilde\o_d$ are in general obviously only defined up to adding either a derivative of a $(d-1)$-form or a gauge-invariant $d$-form, since the corresponding $(d+1)$-form $\o_{d+1}$ then belongs to the same cohomology equivalence class. In case of the WZ forms $\o_{d+1}$ with even $d$, there is, however, an additional ambiguity due to the appearance of the Chern-Simons form. The latter is only defined by the second relations in eqs.~\eqref{defCS3} and \eqref{defCS5}, and can thus be changed by adding a derivative of any form depending on the gauge fields only. Such a modification accordingly affects the expressions for $\tilde{\o}_{2A}$~\eqref{om2A} and $\tilde{\o}_{4A}$~\eqref{om4A}, and thus also the form of the anomaly. The particular choice of the CS forms in eqs.~\eqref{defCS3} and \eqref{defCS5} gives rise to the anomalies in eqs.~\eqref{anomaly2} and \eqref{anomaly4}, which satisfy the WZ consistency condition and are usually referred to as the consistent anomalies~\cite{Bertlmann:1996xk,Bilal:2008qx}.

The above-mentioned ambiguity naturally disappears when only unbroken generators are gauged and the odd-degree WZ forms are closed without adding the CS form. In this case, the anomaly is absent.


\subsection{Overall normalization of the WZ terms}
\label{subsec:normalization}

As is common in effective field theory, the symmetry-based differential-geometric methods employed here fix the nonlinear dependence of the WZ terms on the NG fields, but leave a set of low-energy couplings a priori undetermined. These have to be fixed either by experiment or by matching the effective theory to the underlying microscopic theory. In the case of most physical interest, that is, the WZ term $\tilde\o_{4}$ with a simple symmetry group $G$, this amounts to fixing a single normalization factor. This can be done conveniently by evaluating the contribution of a single selected operator that couples anomalously the NG fields to the gauge fields for the group $G$, such as the operator responsible for the two-photon decay of the neutral pion in QCD. Alternatively, one can fix the overall normalization by evaluating the anomaly functional in terms of the gauge field alone and using eq.~\eqref{anomaly4}. Finally, it is in principle possible to evaluate the whole WZ term directly from the microscopic theory, either by solving the anomalous Ward identities (see e.g.~ref.~\cite{Kaiser:2000ck}), by a direct computation of the anomalous part of the determinant of the Dirac operator~\cite{Salcedo:2000hx}, or by dimensional deconstruction of a higher-dimensional gauge theory~\cite{Hill:2004uc,Hill:2006wu}.


\section*{Acknowledgements}

We would like to thank Joe Davighi and Andreas Wirzba for insightful discussions on Wess-Zumino terms. This work has been supported by the grant no.~PR-10614 within the ToppForsk-UiS program of the University of Stavanger and the University Fund.


\bibliographystyle{elsarticle-num}

\end{document}